# Data selves and identity theft in the age of AI

Tim Gorichanaz, Drexel University

## Introduction

Hundreds of trucks blockaded several border crossings between the United States and Canada as well as the streets of Ottawa, the Canadian capital. Thousands of pedestrian protestors joined in. This "Freedom Convoy" went on for about three weeks in early 2022, stretching an already threadbare supply chain and stoking political divisions in several countries. The Freedom Convoy caused economic losses of at least $700 million, with some estimates as high as $6 billion. Ostensibly, it all began as a truckers' protest against Covid-19 vaccination requirements for crossing the border and then grew into a general protest against pandemic restrictions and mandates. Yet investigative journalists have discovered that the Freedom Convoy movement was sparked not by grassroots dissent but by anonymous actors running Facebook groups through a stolen account.

According to *Grid*, the Facebook account of a Missouri woman was used to launch several Facebook groups from January 26 to 28, accruing over 340,000 members and 7,500 posts. When *Grid* spoke with the woman, "she said her account was hacked and she was not involved with the groups. 'Someone stole my identity on Facebook,' she said. 'I don't know how.'"[1]

Today nearly 3 billion people have active Facebook accounts, some 60% of global internet users. Even if we're not on Facebook, we must manage numerous online accounts; and as we navigate the internet, we unwittingly leave digital traces like flakes of skin. So though we sometimes talk about our "online identity" or even "online identities," the truth is that these accounts and traces are so heterogeneous and diffuse that it is impossible to speak about identity as a distinct object at all. And as we can see in the example of the Freedom Convoy, cunning actors can exploit this heterogeneity and diffuseness to commit crimes of far greater reach and impact than simple credit card fraud.

---

[1] https://www.grid.news/story/misinformation/2022/02/08/the-hacked-account-and-suspicious-donations-behind-the-canadian-trucker-protests/

The line "On the internet, nobody knows you're a dog" first appeared in a 1993 *New Yorker* cartoon, and three decades later it is as true as ever. What has changed, as the internet has become integrated with nearly every aspect of life and society for most of the world's population, is the power and reach of any dog online.

This chapter considers identity theft in the digital age, particularly in the context of emerging artificial intelligence (AI) applications. Because this issue is inextricable from questions of big data and selfhood, we will begin with discussions of those topics. The next section reviews the concepts of data selves and data doubles as well as the process of identification in the digital age. After that, we will review the literature on online identity theft, including its theoretical and empirical aspects. As we will see, AI technology has increased the speed and scale of identity crimes that were already rampant in the online world, even while it has led to new forms of crime detection and prevention. As with any new technology, AI is currently fueling an arms race between criminals and law enforcement, with end users often caught powerless in the middle. To close the chapter, we will explore some emerging directions and future possibilities of identity theft in the age of AI.

## Data and the Self

Life is "a blooming, buzzing confusion," as William James famously put it, and to communicate at all we need to work with concepts and categories. For instance, there are infinite colors on the spectrum, but we only name a handful of them. When it comes to describing our own being, two major concepts are self and identity (Gorichanaz, 2020, pp. 67–87; see also Ahmed, 2020, pp. 24–45). These are abstractions, meaning they necessarily do away with many details for the sake of maintaining a small, portable kernel.

Lately, there has been concern over how digital technologies—and data-driven AI in particular—interact with the concepts of self and identity. Gilbert and Forney (2013) trace the way that selfhood has been shaped in response to emerging technologies for millennia. The pre-modern or *social* self, which defined humanity until the last few centuries, was circumscribed by social role; the modern or *psychological* self was rooted in mental life; and the post-modern or *multiple* self, which emerged during the era of globalized digital technology, is fragmented and contextual. Gilbert and Forney observe that today's emerging technologies, such as AI, cloud computing and immersive virtual worlds, are giving rise to a fourth sense of self, a *distributed* self, in which we conceive our selfhood as spread across multiple devices, platforms and ways of interacting.

We have always been more than just our bodies—we are thinking things, with memories and tools and goals—but our contemporary distributed self is comprised in part by digital data, in many cases stored and owned by others. Many of the digital documents that supplement my memory and guide my work are stored in the cloud, as are the photos on my smartphone that document my memories, activities, and interests. I generate data as I browse the web, move throughout my university and city, make purchases, and communicate with others. Douglas-Jones (2021), in an anthropology essay, seeks to clarify the multiple understandings of data as it intersects with selfhood (data as, data about, data of) and the multiple purposes of data collected and used by different entities. She offers three vignettes that demonstrate these different understandings: allowing companies to assemble a *data double* for one's individual benefit in some area, such as creditworthiness; resisting corporate data collection regimes by assembling one's own *data composite*; and using *data aggregates* to clarify or change relationships between states and people, such as particular minority groups

Of the three types of data self that Douglas-Jones (2021) outlines, the first two have the most relevance for identity theft and will be described here in more detail. In the literature, scholars tend to use the terms *data double* and *data self* for these concepts. Roughly, a *data double* is a data profile constructed by a corporation to, for instance, serve personalized ads; and a *data self* is the collection of data that a person creates about themselves through self-tracking technologies. While data doubles are external, often opaque, and involve little agency or even awareness, data selves are just the opposite. To be sure, the reality is somewhat more complex, and there is some fuzziness between these two categories, but they serve well enough for analytical purposes.

## Data Doubles

If you have ever come across an advertisement online that seemed to be speaking directly to you, then you've brushed with your data double. Data doubles are the profiles assembled by Google/Alphabet, Facebook/Meta, and other companies to represent and predict the behavior of specific users, including shopping and browsing data, demographic characteristics, and more. Just as stunt doubles in Hollywood do not always accurately represent the stars they're supposed to, our data doubles sometimes falter. Perhaps you've come across an ad that didn't quite land. While data doubles are most familiar in the commercial context, they are increasingly used across the digital world.

The data double phenomenon stretches back to the early twentieth century with the rise of the bureaucratic society (Koopman, 2019) and, of course, digital computing (Cheney-Lippold, 2017). An early theorist of the phenomenon was psychologist and philosopher Harré (1984), who wrote of the *file-self*—that version of ourselves comprised of documents, reports and transactions. Later, Solove (2003) wrote of *digital dossiers*. Another term sometimes used for this phenomenon is *digital doppelganger*.

Surveillance is a key issue addressed within data double scholarship. The seminal paper in this vein is "The Surveillant Assemblage," by Haggerty and Ericson (2000). In this article, Haggerty and Ericson explore the implications of the then-emergent digital database technologies on human identity and autonomy. The authors describe how, as discrete surveillance systems begin to converge, people's data doubles come to have their own social lives and materiality distinct from the flesh-and-blood people from which they were assembled. In the name of shopping rewards, crime prevention, public safety, etc., people are abstracted into data flows, which are then reassembled in different locations. In this early work, Haggerty and Ericson predicted that this would transform the purposes of surveillance and the meaning of privacy. That prediction was borne out, as Zuboff (2019) documents in *The Age of Surveillance Capitalism*. In this book, Zuboff describes an emergent form of capitalism based on the dynamics observed by Haggerty and Ericson. Today, companies collect and sell data as "behavioral futures" on predictive markets. This drives the datafication of more and more aspects of human experience, and the predictive markets then not only predict our behavior but also proleptically modify it. In this context, Zuboff foresees dire consequences for democracy and freedom.

In addition to theory, a number of studies have examined people's attitudes and experiences of encountering and negotiating their data doubles. Lupton (2021), for example, presents a wide-ranging interview study of Australian adults' awareness and perceptions of data surveillance and datafication. She found that people were aware of the issues but did not share the dystopian sentiments common to authors such as Zuboff (2019). People thought the data comprising their data double was superficial, while they still had deeper aspects of themselves that were closed off from datafication. They found their online selves to be not only incomplete, but also performative in many cases. Speaking to the incompleteness and performativity of data doubles, Kear (2017) presents an ethnography of how people "game" the credit scoring system through Lending Circles, which help marginalized people build their credit. One tension that emerges in this work is the question of whether Lending Circles are unfairly manipulating the financial system or correcting a prior injustice. In another context, Jones et al. (2020) examine students' perspectives on learning analytics at colleges and universities in the United States. Like in the world of finance, students find that their institutions' data collection practices are largely opaque to them and there is no informed consent involved; in contrast to financial institutions, though, students have a higher sense of trust in their school because of its nonprofit status—trust that would evaporate if their data were to be used commercially.

As this research suggests, data doubles are characterized by opacity and a lack of choice or agency. Resistance and opting out are possible in some contexts, such as leaving a given social network or deciding to use one search engine over another. Other contexts, such as the financial system, offer no meaningful choice. For the most part, some form of digital surveillance is the cost for participating in society, and we may do what we must to navigate the ethical gray areas that the surveillant assemblage and surveillance capitalism present.

## Data Selves

While our data doubles are external and opaque, we may also collect and use data about ourselves because and in ways that we want to. This phenomenon has come to be called the *data self*. In an expansive look at the topic, Rettberg (2014) examines blog-writing, selfie-making and wearable self-tracking as manifestations of the relationship between selfhood and technology. She demonstrates how people develop emotional ties to their data and the feeling that seeing your data is a way of seeing yourself. Key to the picture is agency: in these practices, we can edit, shape, curate and share our data in ways that we want to. To be sure, this is not only an individual process, but also a social one; and this point is developed in detail by Lupton (2020) in a series of studies on self-tracking cultures. In this work, Lupton covers questions of governing the self, the entanglement of human bodies and technology, how we value data, and how self-tracking relates to social inequality.

Just like data doubles, data selves have a long history—in this case stretching back millennia in the form of journals and letters. Humphreys (2018) in *The Qualified Self* offers a historical throughline from centuries-old almanacs, pocket calendars and diaries to today's Fitbits and social media. As a scholar of communication, Humphreys is concerned with how any era's practices of self-documentation help us understand ourselves and each other.

Not surprisingly, then, data selves also make use of experimental and emerging technologies. A good showcase of this is Robards et al.'s (2019) work presented on a panel at the Association of Internet Researchers conference, which offers a kaleidoscope of research reflecting on the notion of "data-selfie." These authors explore how data becomes an extension of the self, the consequences of datafication, the political and commercial aspects of data selves, data anxiety among older internet users, and feelings of control in data generation and use. This work, as well as that of Lupton (2020), helps us reflect on the fuzzy boundaries between data doubles and data selves—sometimes the sense of choice and control we may experience with our data selves is only an illusion.

## The Self and Reidentification

Above we discussed the creation of what can be called personal data or personally identifying information. This kind of data comes from a person and is stored in databases. But let us consider the other direction: If what you have is data, how can you trace it reliably back to the right person? This is a question about identification and reidentification. For example, fingerprints are unique and can be compared to identify a person. Photographs on identification cards, too, are fairly reliable. But in the realm of textual and numerical data, things are much less clear. If someone can recite a social security number (SSN), it doesn't necessarily mean that is their SSN—hence banks may look to other data or ask additional questions to identify a person. A corpus of browsing data may seem to suggest a person, but this could be fallible. Moreover, the more data and systems involved, the more complex the picture of personal identification is.

Brensinger and Eyal (2021) provide a framework for understanding the complex processes of personal identification in the age of AI and big data. Their work incorporates insights from sociologists such as Durkheim, Goffman, Foucault, and Deleuze, and their primary purpose is to assist sociologists researching issues of identification, such as identity theft, to better understand where in the system this or that has gone awry. Brensinger and Eyal's framework has three major elements: object, agency, and technique. The *object*, or object of identification, is the person being identified. However, it is not the flesh-and-blood person, but only a partial representation of that person created through disembedding, standardizing and reembedding. Disembedding involves the processes of abstracting and datafication; in reembedding, people shape their behavior to maintain the integrity of the system. Next, *agency* represents questions of who has access to what identifying information, and who cooperates with whom. In identification systems, agency is asymmetrical and complex, involving the person themselves and various bureaucratic entities that play different roles in constraining or affording action. Finally, there are numerous techniques at play in *identification*. These include, for instance, matching data points, establishing baselines for comparison (e.g., government stored photo, signature, etc.), expert judgment, and interactivity (resistance, adaptation).

This system can be exploited through identity crimes such as identity theft, which is the topic of the next section of this chapter.

## Identify Theft

There are myriad cyberattack concerns in the digital age: malware, denial-of-service, password cracking… With AI technology in the mix, there is the potential for these to occur faster, perhaps autonomously, and in new ways altogether. King et al. (2020) provide a comprehensive review of AI crime, identifying a range of AI crimes, some of which have already been perpetrated while others are still only potential. Of particular note, AI introduces thorny questions of responsibility and liability, given that AI systems can be opaque, unpredictable and distributed. King et al. also discuss possible solutions that are emerging in the literature, such as introducing new legal structures and using AI to monitor for AI crime.

Rather than exploring every type of cybercrime, here we'll focus on identity theft. In brief, identity theft is when one party obtains somebody's personal data unlawfully, typically then selling that data or using it to commit fraud. For example, an identity thief may illegally obtain a person's SSN, list of past addresses, and family members' names (identity theft) and then use these details to open a new credit account (identity fraud). Identity theft is not a new crime, of course, but as the internet became more and more central in people's lives around the turn of the millennium, incidents and fears of identity theft rose to the level of a moral panic (Anderson et al., 2008). In the United States, as in many countries, identity theft is the fastest-growing type of consumer crime; about three-quarters of the U.S. population believe they will fall victim to it (Hille et al., 2015). Digital technology has also given rise to online marketplaces for consumer personal data on what has come to be called the dark web (Anderson et al., 2008).

A succinct overview of identity theft for technologists is offered by Aïmeur and Schőnfeld (2011). The authors delineate several types of identity theft, the two most common of which are financial (e.g., related to tax and credit fraud) and medical (e.g., billing treatment to someone else's health insurance account). Aïmeur and Schőnfeld outline several types of information that can be involved in identity theft, including: personal information (name, age, gender, etc.); buying patterns; physical and digital navigation habits; lifestyle information; records of employment, medicine and criminality; and biological information. As Aïmeur and Schőnfeld write, this information can be physical and/or digital, and it can be retrieved through traditional methods such as dumpster diving and social engineering as well as online methods such as phishing and packet sniffing.

In an economics article, Anderson et al. (2008) chronicle the history and societal impacts of identity theft, providing an expansive perspective on the scope of the problem and the issues involved. As these authors write, the shift to an online economy, its attendant rise in anonymous transactions and our new reliance on personal information (e.g., needing a credit score to secure housing and a job) has allowed explosive growth in identity theft and fraud. Anderson et al. review both the direct and indirect costs of identity theft, government responses, institutions available for preventing and detecting ID theft, and policy issues (liability, consumer data protection, credit freezing, and penalties and resources for enforcement).

Perhaps the most notable institution in this area is the Identity Theft Resource Center, founded in 1999.[2] The ITRC publishes research on identity theft as well as provides support for victims of identity theft. Fellow researchers working in this area will find value in their research publications on topics including the emotional and psychological impact of identity theft, the aftermath of data breaches for victims, the intersection of social media and identity theft, cyberattack trends and more.

This section reviews the literature on identity theft in the digital age. We will begin by discussing the emerging dynamics that AI technologies are adding to the picture of identity theft. Then we will turn to privacy, a major concept invoked in the identity theft literature, and then to empirical work on awareness and victimization. Following that, we will examine preventative measures and solutions in the literature to address the problem of identity theft. We will close with some reflections on future developments in this area.

## AI and Identity Theft

Simply put, the proliferation of AI increases the threat of identity theft. At heart, AI relies on data collection and storage, and stored data is subject to being breached. In this context, the practice of data brokering (trading and collecting consumer data from various sources) is a key vulnerability for identity theft (Aïmeur & Schőnfeld, 2011). Over the past several years, data breaches have been on the rise, though one study did find that in 2021 data breaches decreased worldwide—with the exception of the United States, where they increased by 10% over 2020 levels (RiskBased Security & Flashpoint, 2022). According to this same report, the most commonly targeted data are names, addresses and social security numbers.

Besides the weak point of data storage, AI also changes the landscape of identity theft in other ways. First, it introduces new forms of personal information; for example, predictive AI models of users can be considered personal information (Aïmeur & Schőnfeld, 2011). As well, AI potentiates new technological forms of identity theft. For example, AI can allow system penetration without human involvement. Attacks can become automated, faster, and less detectable. One AI-based technique in this arena is neural fuzzing, in which an attack floods a system with a large amount of random input to quickly probe its vulnerabilities. And third, AI introduces new possibilities for social engineering, a "low-tech" approach to identity theft. The ITRC reported the first case of AI-based social engineering in 2019, in which criminals created a deceptive synthetic audio clip (referred to as a deepfake) of a company's CEO asking for money to be transferred to a particular account.[3] This is an example of "vishing" or "voice phishing," which has been attested since 2008 as a means of social engineering (Griffin & Rackley, 2008), but with deepfake technology the method grows more pernicious.

---

[2] https://www.idtheftcenter.org

But AI does not only further enable criminals when it comes to identity theft; it is a double-edged sword. AI technologies can be used by criminals to facilitate theft and fraud, and it can be used by other entities to detect and prevent these. For example, banks and creditors are using AI to automatically detect unusual account activity. If Alice usually shops at certain stores in a certain area with a given frequency, and suddenly her account registers a purchase that doesn't fit the pattern, the system will flag her purchase for human review or automatically block the transaction until she approves it. AI developments in this space, then, are something of an arms race between criminals on one hand and cybersecurity professionals and law enforcement on the other.

## Privacy, Architecture, and Identity Theft

Privacy is a key concept invoked in discussions of identity theft. Some thinkers have conceptualized privacy in individual terms, while others see privacy as an environmental or contextual issue. An example of the former is the seminal paper "A Right to Privacy" by Warren and Brandeis (1890, as cited in Solove, 2003), in which privacy is defined as a "right to be let alone" and "right to one's personality."  More recent conceptualizations of privacy focus on the appropriateness of various information flows within an environment. An example of this approach is found in Solove's (2003) legal scholarship. Identity theft is a clear violation to both approaches to privacy, but the latter approach more comprehensively diagnoses the issue and points toward viable solutions. Here we will consider Solove's proposals in some detail.

In an extended argument, Solove (2003) makes the case that while some privacy violations may be isolated incidents on individuals (e.g., snooping on a neighbor), others are architectural issues—and identity theft is an architectural issue. He means "architecture" here broadly as the ways that our structures shape society in terms of behavior, attitudes, thoughts and interactions. This includes the shape of our physical buildings, but also our other physical and social spaces, as well as digital ones. Architecture can enhance or diminish privacy; consider the effects on privacy of a building made completely of glass compared to one with no windows at all. When viewing privacy as an architectural issue, the flaws that enable identity theft are visible (e.g., what is made possible simply by having someone's SSN). But the law has been ignoring this flaw, focusing instead on isolated, individualized solutions. Such solutions mean that individuals need to have the knowledge, power, and resources to assert their rights; and the most powerless people in society are both the least able to do so and the most vulnerable to identity theft. Other problems with viewing identity theft and privacy as individual issues include a lack of government resources and the fact that individuals may not be aware they have been violated until long afterwards. Solove suggests that architectural issues such as privacy can only be fixed by architectural solutions. Individual rights and assertions are a piece of that, not the whole picture. In the end, he proposes a new architecture for privacy built around Fair Information Practices (which have been in discussion since the 1970s; see Ahmed, 2020, pp. 634–649). This would make it possible for people to discover what information about them is collected and held by third parties and how it is being used, to correct or amend such records, etc. In short, this proposal would place the burden for personal data protection not on the individual victims but on the entities storing and using that data.

---

[3] https://www.idtheftcenter.org/post/first-ever-ai-fraud-case-steals-money-by-impersonating-ceo/

Also along these lines, Wang et al. (2006) present a contextual framework describing identity theft. Their framework shows the various stakeholders involved in identity theft—the identity owner, the identity issuer, the identity checker, the identity protector, and the identity thief—as well as the interactions among these. Their framework is meant to help analysts understand how it happens, develop security solutions, and assess the strengths and weaknesses of proposed fixes. Ahmed (2020) provides a broader framework, one showing all the possible forms of identity crime and how they relate to each other, modeling the various entities and actions associated with such crimes (e.g., production, acquisition, transfer, etc.).

## Attitudes and Awareness

A body of empirical research has examined people's awareness of and attitudes toward identity theft. By and large, people are quite aware of the risk (if only abstractly), and the predominant attitude appearing in the literature is fear. To date, most of this research has been done in the United States, most often with students (Choi et al., 2021). This section reviews some of this work in effort to offer a comprehensive but succinct global snapshot.

Though most people are aware that identity theft is possible, researchers have sought to learn more about the connections between this knowledge and people's behavior. In a recent study, Khan et al. (2021) examined people's cybersecurity awareness and self-disclosure behavior on social media in a quantitative survey including a range of demographics and frequency of internet access. They found that people were broadly aware of the risks and performed cost-benefit analyses before disclosing information, providing evidence for the effectiveness of cybersecurity education. Cross (2017), in a study of senior citizens, provides a contrasting view, showing that older people may unwittingly expose their information through their online activities, primarily because they have incorrect beliefs about the risk and likelihood of identity crimes. Cross suggests that the prevailing model of individual education and self-regulation falls short for some user groups, such as elderly people.

Because identity theft is closely linked to data security, researchers on identity theft should also consider attitudes and behaviors regarding data breaches (which are, unfortunately, frequently in the news). In a report from the RAND Corporation, Ablon et al. (2016) investigated people's attitudes and responses to data breach notifications. They found that 43% of U.S. adults were notified about a data breach involving their data, with about half of those notifications occurring in the past year. About 44% of people learned of the breach from the media before receiving an official notification. Afterwards, 64% made use of the free credit monitoring offered, and customer attrition was only 11%. Their research suggests that customers are most satisfied when the breach notification is timely, and when they are kept up to date on improved security measures and offered identity protection and credit monitoring services. Ablon et al. also consider the prospect of government-mandated breach disclosure laws, discussing the complex pros and cons. As of early 2022, there is still no such U.S. law in effect. In contrast, the European Union does have such laws in effect, as do Australia and China.[4] For an in-depth review of identity crime legislation in the United States, Canada, Australia, and the United Kingdom, see Ahmed (2020, pp. 252–542).

Much of the work on people's attitudes and awareness examines fear. Indeed, fear of crime has long been of interest to criminologists; this is because fear of crime is generally far more widespread than actual incidents, and such fear can have detrimental emotional and physical impacts (Henson et al., 2013). Roberts et al. (2013) examine fear of cybercrime specifically, finding that it can be weakly predicted by fear of crime in general and specific internet exposure. But given the weak prediction power, more research is necessary; cybercrime has not yet seen as much attention as offline crimes. In more recent work, Choi et al. (2021), using data from a South Korea crime victimization survey, find that fear of identity theft is higher than of other types of crime, and that fear of being victimized was related to online exposure (banking and shopping etc.) and exposure to motivated offenders (getting phishing emails).

Though many of us are used to working online, this is not the case worldwide. Jordan et al. (2018) and Jibril et al. (2020) provide insights from Slovenia and Ghana, respectively, societies that are currently transitioning to online economies. Both studies found that fears of financial loss and reputational damage related to people's hesitancy to engage in online financial transactions.

## Aftermath for Victims of Identity Theft

In addition to people's attitudes and behaviors regarding identity theft, there has been a small body of research looking at people's responses to becoming a victim of identity theft. Golladay and Holtfreter (2017), using data from the U.S. Bureau of Justice Statistics, showed that, in addition to financial and time losses, victims of identity theft also experienced emotional (e.g., depression) and physical (e.g., poor health) symptoms. In a doctoral dissertation, Gideon (2020) renders this in much greater detail, through phenomenological interviews with victims of identity theft. Gideon found that victims experienced fear, hopelessness, apathy, reactivity, and anger, and that the mixes of these emotions differed for each victim. In terms of behavioral change after victimization, Gideon found that fear drove the most change, while apathy did not drive change.

In another study, Gies et al. (2021) looked specifically at identity theft victims who made use of the ITRC. They found that victims who contacted the ITRC may have had more severe cases than those who did not contact the ITRC. Specifically, victims who contacted the ITRC reported more personal problems with finances, employment, family, friends, and physical health one year after their identity theft incident. They also reported longer remediation time and lower degrees of full problem resolution. However, these participants reported significantly fewer mental health problems than those who did not contact the ITRC. All in all, Gies et al.'s findings show the myriad, complex and long-lasting problems that victims of identity theft experience, as well as the ways that existing services such as the ITRC may be helpful.

---

[4] https://en.wikipedia.org/wiki/Security_breach_notification_laws

## Preventing Identity Theft

As mentioned above, the prevailing methods for preventing identity theft, and responding to identity theft once it occurs, place the onus on individuals. For example, guidance appearing across the public and trade literature suggests that people should: protect their account credentials; use temporary emails and throwaway accounts where prudent; and monitor their online identity using services such as Identity Guard, which includes dark web monitoring. Like Khan et al. (2021) discussed above, Lai et al. (2012) found that educating the public on such tactics does reduce risk of identity theft. Specifically, Lai et al. investigate the effectiveness of both conventional coping methods (e.g., shredding bank statements, monitoring credit history) and technological coping methods (e.g., using firewalls, keeping software up to date), finding that both are effective in reducing identity theft risk.

However, some work calls into question the ultimate effectiveness of such strategies. In a theoretical argument, Whitson and Haggerty (2008) find that leaving identity theft as an individual problem, which they call an ethos of "care of the virtual self," is ultimately not practicable. This ethos fits into a broader orientation toward individual self-care in the digital age, which asks each person to care for their data double. And while this may not be problematic in some contexts, it is not sufficient to prevent identity theft. This echoes Solove's (2003) description of identity theft as an architectural problem rather than an individual one. Providing empirical grounding to these claims, Shillair et al.'s (2015) research points out that people are not taking the necessary precautionary steps to protect themselves, even when they know about these steps.

Most recently, Piquero et al. (2021) conducted a focus group with cybersecurity professionals on preventing identity crimes. Their work reports on these professionals' assessments of numerous technological solutions, including various scanning and monitoring programs, as well as their observations about the current risk environment. The vast majority of the technological solutions currently on offer are perceived by these experts as ineffective (e.g., prevention software, end-user agreements, knowledge-based authentication). On the other hand, the experts view chip-enabled credit/debit cards as highly effective, and the following as partially effective: two-factor authentication, biometric authentication, and AI-based authentication. Beyond these solutions, there are a number of emerging AI-based techniques for detecting and preventing identity theft and other cybercrimes. Al-Khater et al. (2020) provide a comprehensive review of such techniques, including statistical, machine learning, data mining, computer vision, biometrics, cryptography and forensics approaches. In perhaps the most comprehensive treatment to date, Ahmed (2020, pp. 543–600) outlines dozens of strategies that can be used to prevent and minimize identity crimes at all levels, from individuals to businesses to government.

All this suggests that, while some emerging technological solutions will go some way in preventing identity theft, there is more that could be done to address this issue architecturally. Building on the arguments of scholars such as Solove (2003), Van Loo (2019) points to "the missing regulatory state" in a law review article arguing for the regulation of digital platforms such as Twitter and Facebook and tech companies such as Amazon and Alphabet. Van Loo argues that a too-narrow understanding of privacy is often used by these entities as a pretext for deregulation; he paints a picture of how regulatory monitoring is indeed possible without having the negative aspects typically associated with government surveillance. In short, regulators need access to certain data held by tech companies and digital platforms to ensure consumer safety and company compliance, and at present they do not have such access.

## The Future of Data Selves and Identity Theft

As we move further into the digital age, we are carrying more devices with us, using the internet more frequently and for more activities, and thus rendering more of our experiences as data. As AI is used in more of our devices and systems, we benefit from their increased flexibility, efficiency, and convenience; yet this technology also increases the possibilities for other entities, from corporations to governments, to shape our movements. It also holds possibilities for new forms of self-knowledge and reflection, depending how technology and society develop.

To close, I want to shine a light on two emerging technologies in this space that researchers should explore further: blockchain and synthetic identity fraud. Blockchain is a technology based on a shared, public ledger that makes data exchanges transparent to all relevant parties, facilitating trust and communication. Perhaps the most famous application of blockchain is in new forms of currency such as Bitcoin, but blockchain can also be used on smaller scales to track orders, accounts, payments, manufacturing, etc. Blockchain could also be used to register human identities in ways that make certain forms of identity theft impossible. To this end, Rana et al. (2019) provide an overview of blockchain-based identity solutions. Further, there is significant potential in the combination of blockchain and AI. There is nascent work in this direction from entities such as IBM. The combination of blockchain and AI can improve trust and transparency of AI decisions, speed up blockchain transactions, etc.

Next, AI technologies may shift the goalpost from identity theft to synthetic identity fraud (Walker-Moore, 2018). Whereas in identity theft, criminals impersonate an existing person, with synthetic identity fraud, a false identity is created. Using a synthetic identity, fraudsters can, for example, open and use a line of credit for a person who does not exist. Such fraudsters may take advantage of weak verifications, knowledge about the social security number registration process, etc., as well as emerging AI technologies such as synthetic media (deepfakes) that would allow criminals to create photos, video and voice of people who do not exist. Walker-Moore (2018) provides a history and prognosis of this problem, as well as directions for solutions.

Identity theft is a danger that has been growing since the dawn of the digital age, and it is not likely to go away. New technologies, including AI, introduce new vectors for identity criminals to exploit. At present, virtual assistants, smart speakers, and customer service chatbots present new opportunities for criminals. At the same time, these technologies also allow others to more easily detect and prevent identity theft, leading to an arms race. Whether the arms race can end remains to be seen; likely it would take more than just technological innovation, but also new approaches to architecture and regulation as advocated by scholars such as Solove (2003) and Van Loo (2019).